# What is the Contribution of Intra-household Inequality to Overall Income Inequality?
# Evidence from Global Data, 1973-2013

Deepak Malghan
Indian Institute of Management Bangalore
dmalghan@iimb.ernet.in

Hema Swaminathan[1]
Indian Institute of Management Bangalore
hema.swaminathan@iimb.ernet.in

August 2016

**Abstract:**
Intra-household inequality continues to remain a neglected corner despite renewed focus on income and wealth inequality. Using the LIS micro data, we present evidence that this neglect is equivalent to ignoring up to a third of total inequality. For a wide range of countries and over four decades, we show that at least 30 per cent of total inequality is attributable to inequality within the household. Using a simple normative measure of inequality, we comment on the welfare implications of these trends.

---

[1] Acknowledgements: We would like to thank Prachi Yogesh Jain and Akilesh Lakshminarayanan for excellent research assistance, and Suchitra JY for many insightful discussions on this topic. All errors remain our own.

# 1. Introduction

Intra-household inequality remains a neglected area of research in the current inequality discourse. With globalization affecting all corners of the world, global inequality is *cause celebre*; yet domestic inequality concerns cannot be negated. "Because the world is not united under a single government, however, we cannot dispense with the need to look at individual nation-states." (Milanovic, 2016, pp1). Within national boundaries, many group-based inequalities are studied depending on the country context – race, religion, ethnicity, caste are often the relevant lens to sharpen the focus on inequality (Kanbur, 2016). Surprisingly, a study of the household as a social unit where inequalities play out has largely been missing in this literature. A few exceptions to this overall neglect include (Rodriguez, 2016), Malghan & Swaminathan(2015), Lise & Seitz (2011), Sahn & Younger (2009) and Haddad & Kanbur (1990).

The intra-household resource allocation literature has contributed tremendously to our understanding of gender inequalities within the household. There is robust empirical evidence that there is heterogeneity within the household in terms of resource allocation between men and women. Generally, women are less likely to earn the same level of income as men, less likely to own key assets and consequently, less likely to be as wealthy as men. There are of course, spatial and temporal variations in the levels and extent of such inequalities reflecting evolving socio-cultural norms, structural changes in the economy, and policy interventions. These within-household variations can provide insights into the long-term trends of inequality; in fact ignoring the household dynamics can lead to a flawed understanding of overall inequality patterns (Chiappori & Meghir, 2014).

A key problem with measuring poverty or inequality is the disconnect between the unit of analysis and the unit of data collection. Typically, one is concerned with the wellbeing of the individual, but the smallest unit for which data is collected is usually the household. This is mostly true for income, consumption or wealth data, which are the typical focus of inequality analysis. To move from the household to the individual, a per capita method is adopted that entails equally apportioning the household total amongst all its members. Sometimes, equivalence scales are used to adjust these figures for age and sex. The problem with this methodology is that it assumes away any intra-household inequality with the consequence that one gets an underestimate of poverty and inequality (Haddad & Kanbur, 1990; Lahoti, Suchitra, & Swaminathan, 2011; Vijaya, Lahoti, & Swaminathan, 2014).

A greater challenge arises from the fact that within a household many types of goods are produced and consumed. An unambiguous normative welfare interpretation of intra-household income inequality is not possible given that income could be differentially applied between public goods that everyone in the household can enjoy (housing is the classic example of such a good) and private goods that benefit a single or only a few household member(s) (Chiappori & Meghir, 2014; Klasen, 2004; Malghan & Swaminathan, 2015). Without detailed data on consumption by individual household members, one cannot say much about consumption poverty or inequality for individuals.

The goal of this paper is to present the contribution of intra-household income inequality to overall income inequality while for the present, abstracting from the concerns of how this income may be channelled within the household. This knowledge by itself is important for several reasons. First, there is now enough evidence to suggest that the income pooling assumption does not always hold, *i.e.,* the identity of the income recipient matters and can affect intra-household allocation in the non-material domain.[2] Further, this understanding of within household inequality may provide insights on how to rein in overall inequality.

There is an extensive literature from OECD countries that studies the contribution of women's earnings to household earnings inequalities (for example, Esping-Andersen, 2009; OECD, 2011, 2015). On average, women's earnings have been rising largely to due to reductions in the gender employment gap (difference in employment rates between men and women) and in the gender wage gap. Although labor market conditions have improved overall, women are still more likely to work part time and continue to be segregated in lower paying occupations (OECD, 2015). These studies largely find women's rising earnings tend to reduce household inequalities even as there may be some regional variations (Gregory, 2009; Harkness, 2013). A recent study by Nieuwenhuis, van der Kolk, & Need (2016) that examined the long-term trends (1973 – 2013) across 18 OECD countries reinforced these results. A high spousal correlation between earnings could exacerbate household inequality, but is also countered by the reduction in earnings inequalities within women. However, there is no explicit consideration of earnings inequality within the household in these studies.

This paper aims to fill this gap by documenting the trends in intra-household inequality and its contribution to overall inequality for a broad set of countries. In this paper, overall inequality is decomposed into *within-group* and *between-group* inequalities by treating the household as a group. Thus, we calculate the contribution of inequalities *within the household* (intra-household inequalities) to overall inequality. The analysis focuses on differences in earnings between partnered couples to abstract from concerns of the life-cycle effect of earnings. Within a household, one expects that there will be inequality in earnings across generations due to the effect of age and experience, which need not be the case in a conjugal unit.

---

[2] Analogous to the identity of the income earner being relevant is also the identity of the asset owner being relevant for intra-household dynamics. Even for classic public goods like housing, empirical evidence shows that ownership of the asset makes a difference, particularly for women (*cf.* Section 2). While non-owners may be able to use and enjoy the benefits of the asset, there are some advantages that may accrue to only the owners; ability to use the asset as a collateral, or not having to worry about continued access to the asset in the event the household dissolves.

To the best of our knowledge, this is the first study that decomposes overall income inequality into between household and within household components over a large set of years, countries, and regions. We apply the decomposition to countries in the LIS global database that has income micro data over three decades. The remainder of this paper is organised as follows. The next section provides a brief overview of the intra-household literature with a particular focus on inequality issues. The welfare implications of intra-household inequality are also discussed here. The data and methods are outlined in section 3. The results are discussed in section 4, while the final section concludes.

## 2. Intra-household Inequality and Welfare

The unitary model developed by Becker (1974), was the first model to unpack household behaviour where the family is conceptualised as a single decision making unit with an altruistic decision maker. One of the important assumptions of the unitary model is that of income pooling. Income pooling essentially means that as long as prices and total income remain constant, household behaviour is independent of the identity of the income recipient. However, the collective models of household behaviour not only argue against income pooling, but also emphasize that the redistribution of income among household members "[c]an also influence the relative 'power' of the partners." (Browning, Chiappori, & Weiss, 2014, pp 3).

Empirical evidence supports the notion that the identity of the income recipient is germane for household decision-making processes as well as welfare outcomes. When women control resources within the household (either in terms of income, wealth or through transfers), it has intergenerational positive impacts via better investments in children's health and education (Allendorf, 2007; Bobonis, 2009; Lundberg et al., 1997; Park, 2007). Greater resource control or ownership of assets is also beneficial for women's status, power, and wellbeing. It leads to greater empowerment as measured by involvement in household decision-making or mobility (Anderson, & Eswaran, 2009; Swaminathan, Lahoti, & Suchitra, 2012b), reduced risk of experiencing intimate partner violence (Bhattacharyya, Bedi, & Chhachhi, 2011; Oduro, Deere, & Catanzarite, 2015; Panda & Agarwal, 2005), and risk of engaging in transactional sex for unpartnered women (Muchomba, Wang, & Agosta, 2014).

Intra-household inequalities are experienced across material and non-material dimensions. Studies have documented disparities with respect to investments in education and health, nutrition, ownership of key assets, wealth levels and consumption expenditures (Lise & Seitz, 2011; Sahn & Younger, 2009; Swaminathan, Lahoti, & Suchitra, 2012a). Most often though not always, gender is the fault line along which inequalities are starkest, with women and girls discriminated relative to men and boys.

Haddad & Kanbur (1990) in their examination of nutritional status in Philippines highlight the perils of ignoring within household distribution. Their findings show that errors of a magnitude of at least 30% are made in the levels of poverty and inequality when intra-household distributions are neglected. However, there is hardly any impact on the poverty or inequality ranking of different socio-economic groups when intra-household allocations are accounted for. More recently, using sex-disaggregated data on asset ownership and wealth in Karnataka, India, Vijaya et al. (2014) find substantial differences between men and women (almost 34 percentage points) when individual information is used to calculate multidimensional poverty rates. When household poverty status is assigned to individuals, there is a difference of only one percentage point in the poverty rates of men and women. Using this same data, Malghan & Swaminathan (2015), show that for coupled households, 32% of the total wealth inequality (as measured by intra-household contribution to overall Thiel index of wealth inequality) is driven by inequality within the household. Klasen & Lahoti (2016) examine individual level multidimensional poverty and inequality using nationally representative data from India. Their findings also point to considerable diversity within the household along the lines of gender and age. The poverty rate is higher for women and older adults when using an individual measure as opposed to a household measure.

Using individual-level body mass index (BMI) for a set of seven countries, Sahn & Younger (2009) find that of the total inequality in BMI at the country level, at least 55% is explained by within household inequality. A recent study examining four indicators of well-being for boys and girls reinforces the importance of evaluating intra-household inequalities to overall inequality (Rodriguez, 2016). Using data from 27 countries, the author shows that the contribution of within-household inequality as well as the direction of gender bias varies by indicator, suggesting wide variation in household dynamics.

Lise & Seitz (2011) use a collective model of household behaviour to estimate consumption inequality for UK over the time frame of 1968-2001. The authors argue that women's labor supply and wage rates had increased substantially during this time and would have an impact on intra-household inequality as well as overall inequality. Their results show that in the early years, neglect of intra-household inequality could underestimate individual consumption inequality by as much as 50%. This was reduced to 25% during 2000 due to greater marital sorting on earnings over time.

Almost all evidence against the income-pooling hypothesis has used a gender lens and advanced our understanding of gender inequalities and relations within the household. In this paper as well, we examine intra-household gender inequality in coupled households. Gender inequality within the household may or may not mirror the trends in country level gender inequality; examining these correlations is in itself an important exercise. The paper is a first attempt to investigate if intra-household inequality can be accounted for in the larger discourse of inequality. The relationship between these variables is complex and depends on several factors and will involve detailed country level analysis. Nieuwenhuis et al. (2016) show that increases in household earnings inequality is positively associated with correlations between spouses' incomes. This is assortative mating when men and women with similar education profiles form a household. If they are both working with the same intensity, then their incomes are likely to converge.

## 2.1 Aggregate Household Welfare

Even as it is incommensurable to make welfare comparisons across households based on intra-household inequalities due to the public goods issues discussed above, it possible to make comparisons regarding welfare loss.

Consider household *i* with average net personal income $\bar{Y}^i$, and an intra-household income distribution $\Phi^i$:

$$W_j^i = U_j^i(\bar{Y}^i, \Phi^i) \qquad [1]$$

$W_j^i$ is the aggregate household welfare evaluated by individual *j* in household *i*. It is important to note that aggregate household welfare evaluated by some other person, $k \neq j$ can be different from one evaluated by *j*. In the subsample of heterosexual coupled households, this allows for the household welfare function of man to be different from that of the woman. Let $\widetilde{W}_j^i$ be the maximum welfare this household can achieve with perfect intra-household income equality ($\widetilde{\Phi}$).

$$\widetilde{W}_j^i = U_j^i(\bar{Y}^i, \widetilde{\Phi}) \qquad [2]$$

As measured by individual j, the welfare lost due to intra-household inequality is:

$$\Delta_j^i = 1 - \frac{W_j^i}{\widetilde{W}_j^i} \qquad [3]$$

With standard egalitarian preferences, $\widetilde{W} \geq W$ so that $0 \leq \Delta \leq 1$ and $\Delta$ simply represents the fraction of aggregate household welfare lost due to intrahousehold income inequality. While welfare is not directly comparable across households, welfare-loss computed by each household (or even separately by individuals within a household) is commensurable across households. $\Delta^i > \Delta^k$ implies that fraction of welfare lost in household *i* is greater than in household *k*, as measured by specific individuals in respective households. This difference can reflect differences in respective intra-household distributions, differences in public and private consumption in the households, or more typically a combination of the two.

## 2.2 Atkinson Intra-household Welfare Loss Metric

We use a simple welfare theoretic framework pioneered by Atkinson (1970) to estimate aggregate welfare effects of persistent intra-household inequality. Let $\Theta_j^i$ be Atkinson's equally distributed equivalent income (EDEI). $\Theta_j^i$ represents the (equivalent equal incomes) for each of the household member such that aggregate household welfare remains unchanged from the one obtained under extant distribution of income (Atkinson, 1970). Let $\Theta_j^i$ be the EDEI for household *i* as evaluated by its member, *j*. Using Eq. (1), and once again denoting perfectly equal distribution by $\widetilde{\Phi}$, we obtain:

$$W_j^i = U_j^i(\bar{Y}^i, \Phi^i) = U_j^i(\Theta_j^i, \widetilde{\Phi}) \qquad [4]$$

EDEI calculated in Eq. (4) enables the calculation of the Atkinson welfare loss metric:

$$\Delta A_j^i = 1 - \left(\frac{\Theta_j^i}{Y^i}\right) \qquad [5]$$

ΔA in Eq. (5) is consistent with the general welfare loss metric Δ defined in Eq. (3). The difference between average income and EDEI ($\Theta_j^i$) represents the intra-household income equality trade-off from the perspective of person *j*, and $\Theta \leq Y$ so that $0 \leq \Delta A \leq 1$. We illustrate the actual computation of the Atkinson metric in Appendix A.

## 3. Data and Empirical Approach

This paper uses data from the Luxembourg Income Study (LIS) Database, (2016). LIS provides harmonized individual-level income micro-data across a range of countries. Initially, LIS data were mainly from high-income countries, which of late have been expanded to include several middle-income countries as well (Gornick & Jantti, 2013).

Given the main focus of this paper is on intra-household issues, the analytical sample is restricted to households where the head is living with a partner (married, or in a consensual union, or co-habiting). Only households where both spouses are between 18 to 65 years of age are retained in the sample.[3] Following previous studies, we consider only heterosexual couples in this analysis (Harkness, 2013; Nieuwenhuis et al., 2016). In total, we have 37 countries; covering the time period from 1973 to 2013 to give us 215 country-time data points and an overall sample of 2,066,800 coupled households (Table 1). The number of data sets per country range from a minimum of 2 (Switzerland, Poland, Paraguay, Georgia and India) to a maximum of 12 (Mexico). For ease of exposition, the results are presented using regional classifications; Asia, South and Central Americas, Western and Neo-Europe[4], Eastern Europe and Middle East. Not surprisingly, Western and Neo European countries are over represented in our sample, largely due to availability of reliable income data.

---

[3] The age restriction is meant to capture the working age population. We realize that a uniform age categorization may not work across countries, especially in those places where self-employment or informal sector employment predominate. These adjustments will be addressed in a future version.
[4] Neo-Europe includes Canada, United Kingdom and the United States of America.

The key variable of interest for this paper is annual earnings, defined as monetary returns to paid employment. For those who are not employed, earnings are set to zero. Thus, households are included irrespective of the employment status of the spouses. LIS data sets are classified as gross or net depending on whether taxes and social security contributions are captured or not. Gross income data was netted down based on household-level tax information or person-level tax information (Nieuwenhuis et al., 2016). Further, data sets classified as mixed (information in the data is a mixture of gross and net earnings) are dropped from the analysis (*ibid*, 2016). Intra-household dynamics are influenced by the actual contributions of each spouse, which is captured by net as opposed to gross earnings (*ibid*, 2016). Negative earnings are set to zero, while the top one percentile were top-coded to the 99th percentile (Harkness, 2013; Nieuwenhuis et al., 2016). Sampling weights are applied in all calculations.

We use the class of Generalised Entropy (GE)[5] measure with $\alpha = 1$ (Theil-T) to calculate the contribution of intra-household inequality to total inequality. The advantage of an entropy index is that it is perfectly sub-group decomposable, unlike the Gini coefficient. In this paper, each household is a group and is comprised of an adult heterosexual couple. The application of a GE measure to intra-household inequality is not typically the norm, but has been used across several studies (Haddad & Kanbur, 1990; Malghan & Swaminathan, 2015; Rodriguez, 2016; Sahn & Younger, 2009).

In this decomposition exercise, inequality due to wealth and non-labor income is ignored. There will certainly be differences in how these are distributed between partners within a household and are also likely to vary depending on the quantile of the distribution. For example, Malghan & Swaminathan (2015) find for Karnataka, India, that there is far greater equality in wealth distribution between spouses in the poorer quintiles (where there is less wealth) than in richer quintiles. However, government transfers could be targeted to women in poorer households, which could be a factor in driving inequality. The extension to include all income sources and wealth is left for future work.

### 4. Results

Globally, overall inequality displays a rising trend, particularly since the 2000s showing a sharp upward trajectory in the latter half of the decade (Figure 1). In the following discussion, we focus on the contribution of within-household inequality to total inequality. We first present the results of the decomposition exercise of net earnings inequality into within-household and between-household shares at the country level (Table 2). These numbers represent country means across the time frame for which data is available. It is noteworthy that, on average, at least one-third of total inequality in the country is due to inequality within the household rather than between households. The within-household share is the dominant contributor to inequality in about 50 per cent of the high-income countries (represented by Western and Neo European region) and makes up almost half of total inequality in another 30 per cent. Fifty per cent of Eastern European countries show a within inequality share of more than 45 per cent. Only Mexico and Uruguay are close to a within-household share of 40 per cent with the rest of the countries showing a contribution between 30 and 35 per cent. Among the few Asian countries in our sample, the range of within-household contribution to total inequality is from 34 to 43 per cent.

---

[5] $E(\alpha) = \frac{1}{n(\alpha^2-\alpha)} \sum_i \left[ \left( \frac{y^i}{\bar{y}} \right)^\alpha - 1 \right]$

Comparisons across countries can be misleading for the simple reason that data availability varies widely across the sample with respect to number of data points and the time period of the data collected. While the average number of data points per country is 6, there are some countries with only two observations (for example, Switzerland, India, Georgia). Globally, female labour force participation was low during the 1970s and 1980s which would result in higher share of within-household inequality. Thus, countries with data only for more recent time periods would look better with respect to the contribution of within household inequality. Switzerland showing the highest share of within inequality is explained because data are available only for 1982 and 1992. In fact, the share of within inequality had fallen sharply by 20 percentage points during that decade in Switzerland (Table A1).

The intra-country within inequality trends suggest that for the most part, the contribution of within household inequality to overall inequality is declining globally, particularly post 2000s (Figure 2 presents the graphs while detailed tables are in the Appendix, Table A1). There are a few exceptions where the within share is rising (Australia, Hungary, Israel, Taiwan for example), but the change is not substantive and is usually seen in the last two data points, so it is not clear that the increase actually represents a trend. The Nordic countries which in general, have a supportive policy environment for women's employment, show declining contribution of within household inequality. Denmark, Finland and Sweden show lower intra-household inequality than Norway or Iceland.

In the Eastern European region, except for Russia and Hungary in the last two years of their time period, the trend is one of declining contribution of within inequality. The Countries in the Central and Southern American regions, with the exception of Mexico and Paraguay have similar number of data points over a decade beginning 2003-2004. At 33 per cent, the mean contribution of within inequality is stable in Brazil, but increased by 2 percentage points in Peru and Uruguay between 2010-2013. Mexico, on the other hand, shows a comparable decline over this period.

In Table 2, we also report the household welfare loss due to intra-household inequality for two plausible values of inequality aversion parameter, $\varepsilon$. Admittedly, the simple picture of welfare loss presented here omits the fact that there could be systematic differences in inequality aversion between countries, across time, and across different social and income groups within a country. Our analysis assigns a single value for inequality aversion ($\varepsilon = 0.25$) or ($\varepsilon = 1$) to all households in our sample. However, the values of $\varepsilon$ that we have chosen are conservative; short of a unitary model assumption, an inequality aversion of 0.25 is easily defended.

Figure 3 presents the trends in women's share of net household earnings for all 37 countries. The lines are LOESS fitted curves (Local Polynomial Regression) and cannot be generated when there are only two data points. Thus, for countries with only two data points (India, Paraguay, Poland, and Switzerland), only a scatter plot is generated. Women's share in net earnings mostly shows an upward trend across all countries, although the maximum share typically does not cross 40 per cent. At almost 47% in 2012, Slovenia is an exception. The lowest shares are seen in 1970s in the developed countries (Germany, United Kingdom and the United States), which is not substantively lower than the share in some countries for relatively recent time frames. For example, in India the share of women's earnings was 18 per cent in 2011 and in Mexico it was 21 per cent in 2012. Women's contributions to earnings seems to have plateaued for countries that are on the right tail of the distribution; Australia, Canada, Denmark, Finland, Israel, Norway, Slovenia, Sweden, United Kingdom, and United States of America.

Figure 4 contrasts two different approaches to examining gender inequality. Panel A presents the average contribution of inequality between men and women within the same household (intra-household inequality) to overall inequality. Panel B presents the popular approach to examining gender inequality where one is interested in inequality between all men and all women. The key difference between these graphs is that in panel A, the group is represented by the household, whereas in panel B, the group is the sex.

Declining trends are observed across both shares, although the start and end points are vastly different. Intra-household gender inequality is significantly higher than inequality between men and women. Focusing on global averages, the contribution of inequality between men and women has declined from 35.9 per cent in 1973 to 6.5 per cent in 2013 representing a 29 percentage point reduction. Over the same time frame, the contribution of intra-household inequality has also seen a decline of 25.7 percentage points, from a mean of 65.9 per cent to 40.2 per cent. Even so, the average contribution of intra-household inequality in 2013 is still greater than the average contribution of inequality between sexes more than four decades ago.

There are several explanations for why earnings inequality between men and women has declined generally; greater participation of women in the labor market, increased hours of work by women, higher wage rates as well as a narrowing of the gender wage gap (OECD, 2015). However, there is no straightforward mapping of how these factors affect inequalities within the household, although the data suggest that both these types of inequalities are moving together (Figure 5). It will greatly depend on the degree of assortative mating prevalent in society; if households are sorted on education or earnings capacity, then intra-household inequality will be low. On the other hand, if there is little marital sorting then intra-household inequality will be high.

Figure 6 shows the relationship between contribution of intra-household inequality and overall inequality. The two panels represent the same relationship using different labels (country and years, respectively). The observed negative relationship between overall inequality and intra-household contribution to overall inequality is likely driven by the particular sub-sample that we have used in our analysis here where the impacts of assortative mating are most salient.

## 5. Conclusion

The preliminary analysis presented here shows that intra-household income inequalities make a significant contribution to overall inequality. Based on LIS micro-data covering 37 countries, we find that at a minimum, within-household income inequality contributes at least 30 per cent of total country-level inequality. The household is the smallest social unit where distributional issues arise. Yet, it is the most ignored in inequality analysis, either due to data or conceptual concerns. Our results suggest that any attempt to address larger inequality concerns cannot afford to overlook within household dynamics. Intra-household inequalities can potentially affect the wellbeing of individuals and households with intergenerational implications.

Further, we also show that inequalities between men and women in the population are not to be confused by inequalities between men and women within the household. Policies that seek to increase women's earnings generally may have to be recast for women in partnered households.

An important caveat is that our results apply only to coupled heterosexual households, which is the obvious unit of analysis in terms of intra-household gender inequalities. If the interest is inequalities within the conjugal unit, this analysis is easily extended to same-sex coupled households. However, household structure – proportion of male and female heads, households with or without dependants – assume great importance in any discussion of national level inequality. An examination of these issues informs an agenda for future work.

**Table 1: Country classification, time period and analytical sample**

| Regions | Start year | End year | No. of datasets | Total no. of households |
|---|---|---|---|---|
| **Western and Neo-Europe** | | | | |
| Australia | 1981 | 2010 | 8 | 43,474 |
| Austria | 1994 | 2004 | 4 | 5,653 |
| Belgium | 1985 | 1997 | 5 | 12,250 |
| Canada | 1981 | 2010 | 10 | 1,23,255 |
| Denmark | 1987 | 2010 | 7 | 1,72,739 |
| Finland | 1987 | 2013 | 8 | 45,641 |
| Germany | 1973 | 2004 | 9 | 1,10,473 |
| Greece | 1995 | 2010 | 5 | 10,897 |
| Iceland | 2004 | 2010 | 3 | 5,411 |
| Ireland | 1994 | 2010 | 6 | 10,820 |
| Italy | 1986 | 2000 | 8 | 31,825 |
| Luxembourg | 1985 | 2013 | 9 | 13,899 |
| Netherlands | 1983 | 2010 | 7 | 26,124 |
| Norway | 1979 | 2010 | 8 | 2,00,471 |
| Spain | 1990 | 2013 | 7 | 41,986 |
| Sweden | 1975 | 2005 | 6 | 40,837 |
| Switzerland | 1982 | 1992 | 2 | 6,402 |
| United Kingdom | 1974 | 2013 | 11 | 78,013 |
| United States | 1974 | 2013 | 11 | 3,09,442 |
| **Eastern Europe** | | | | |
| Czech Republic | 1992 | 2010 | 6 | 32,680 |
| Estonia | 2004 | 2010 | 3 | 5,995 |
| Hungary | 1991 | 2005 | 4 | 3,297 |
| Poland | 1986 | 1992 | 2 | 8,693 |
| Russian Federation | 2004 | 2013 | 3 | 5,757 |
| Serbia | 2006 | 2013 | 3 | 5,997 |
| Slovenia | 1997 | 2012 | 6 | 9,692 |
| Slovakia | 1992 | 2010 | 4 | 14,986 |
| **South and Central Americas** | | | | |
| Brazil | 2006 | 2013 | 4 | 2,54,574 |
| Mexico | 1984 | 2012 | 12 | 1,15,815 |
| Panama | 2007 | 2013 | 3 | 19,022 |
| Paraguay | 2010 | 2013 | 2 | 5,913 |
| Peru | 2004 | 2013 | 4 | 52,494 |
| Uruguay | 2004 | 2013 | 4 | 66,454 |
| **Asia** | | | | |
| Georgia | 2010 | 2013 | 2 | 3,322 |
| India | 2004 | 2011 | 2 | 60,500 |
| Taiwan | 1981 | 2013 | 9 | 86,027 |
| **Middle East** | | | | |
| Israel | 1979 | 2012 | 8 | 25,970 |
| **Total** | | | **215** | **20,66,800** |

**Table 2: Women's share of net household earnings, contribution of within-household inequality and welfare loss (%)**

| Regions | Mean women's earnings share | Mean contribution of within household | Mean Atkinson welfare loss | |
|---|---|---|---|---|
| | | | e=0.25 | e=1 |
| **Western and Neo-Europe** | | | | |
| Australia | 33.2 | 45.9 | 8.5 | 40.3 |
| Austria | 29.2 | 58.2 | 9.3 | 44.2 |
| Belgium | 26.9 | 54.2 | 8.6 | 41.2 |
| Canada | 35.1 | 49.6 | 8.2 | 38.2 |
| Denmark | 42.3 | 44.3 | 5.8 | 27.1 |
| Finland | 43.1 | 43 | 6.1 | 28.5 |
| Germany | 23.6 | 67.2 | 11.2 | 53.2 |
| Greece | 32.1 | 38.4 | 8.7 | 41.8 |
| Iceland | 38.7 | 54.7 | 5.7 | 25.8 |
| Ireland | 35.2 | 43.7 | 9.3 | 44.1 |
| Italy | 25.4 | 46.8 | 10.2 | 49.2 |
| Luxembourg | 24.1 | 60.4 | 10.1 | 48.2 |
| Netherlands | 25.8 | 56.9 | 9.5 | 44.6 |
| Norway | 37 | 52.5 | 6.6 | 30.2 |
| Spain | 29.3 | 45.5 | 10.0 | 47.6 |
| Sweden | 38.4 | 46.5 | 6.4 | 29.8 |
| Switzerland | 15.8 | 73.8 | 12.7 | 60.3 |
| United Kingdom | 35.2 | 47.4 | 8.7 | 40.9 |
| United States | 33.2 | 57.3 | 9.4 | 44.2 |
| **Eastern Europe** | | | | |
| Czech Republic | 40.1 | 47.6 | 7.5 | 35.7 |
| Estonia | 37.4 | 48.8 | 6.8 | 31.9 |
| Hungary | 37.1 | 43.1 | 8.4 | 39.9 |
| Poland | 33 | 48.3 | 9.8 | 47.1 |
| Russian Federation | 40.6 | 41 | 8.6 | 40.9 |
| Serbia | 34.5 | 34.3 | 8.4 | 40.7 |
| Slovakia | 41.9 | 49.6 | 6.9 | 32.7 |
| Slovenia | 46.9 | 43.9 | 6.7 | 31.7 |
| **South and Central Americas** | | | | |
| Brazil | 33.4 | 33.2 | 9.6 | 46.1 |
| Mexico | 14.8 | 40.2 | 11.9 | 57.3 |
| Panama | 27.4 | 35.5 | 10.9 | 52.3 |
| Paraguay | 26.2 | 33.4 | 8.7 | 42.0 |
| Peru | 21.4 | 32.9 | 9.0 | 43.6 |
| Uruguay | 33.2 | 39.6 | 10.2 | 48.9 |
| **Asia** | | | | |
| Georgia | 33.5 | 33.5 | 9.6 | 45.9 |
| India | 16.8 | 38.4 | 10.2 | 48.8 |
| Taiwan | 26.3 | 43.3 | 9.1 | 43.9 |
| **Middle East** | | | | |
| Israel | 34.1 | 43.8 | 8.7 | 41.4 |

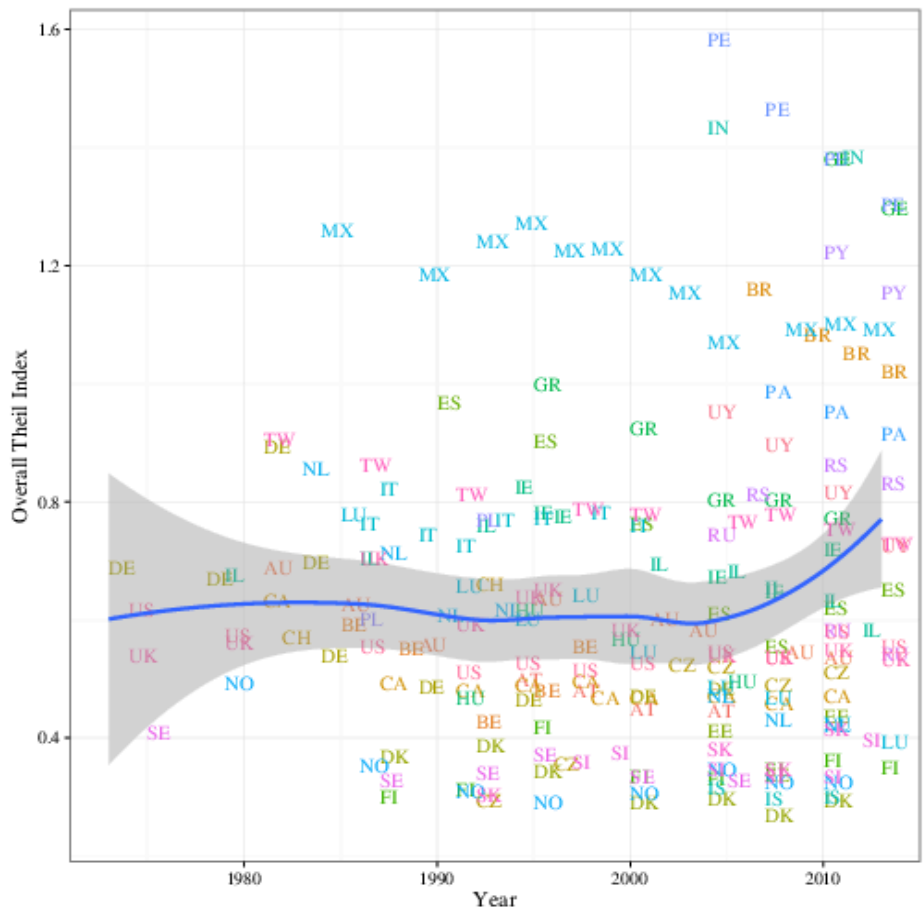

**Figure 1: Trends in overall Theil index, 1973-2013**

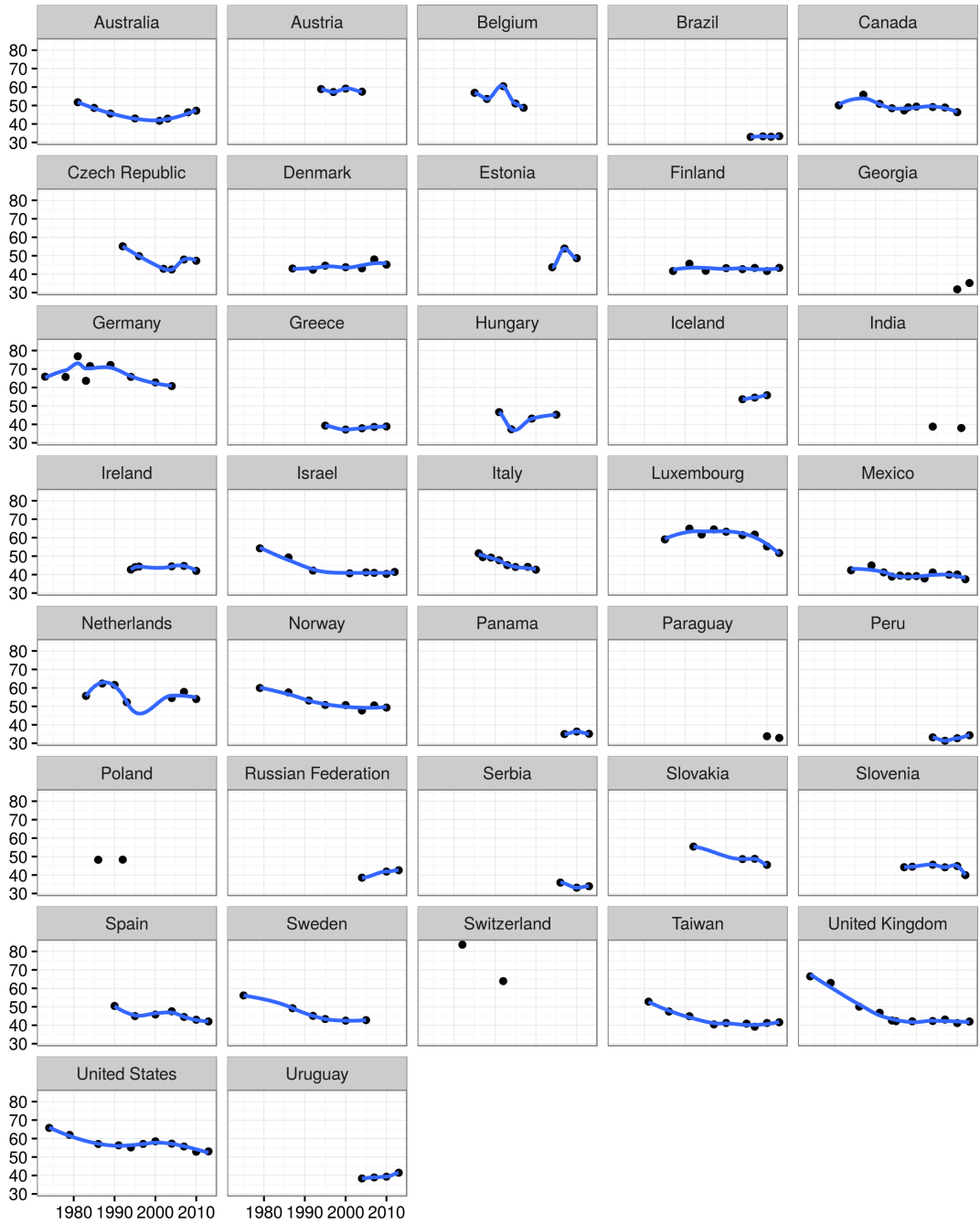

**Figure 2: Contribution of within household inequality to overall inequality (%)**

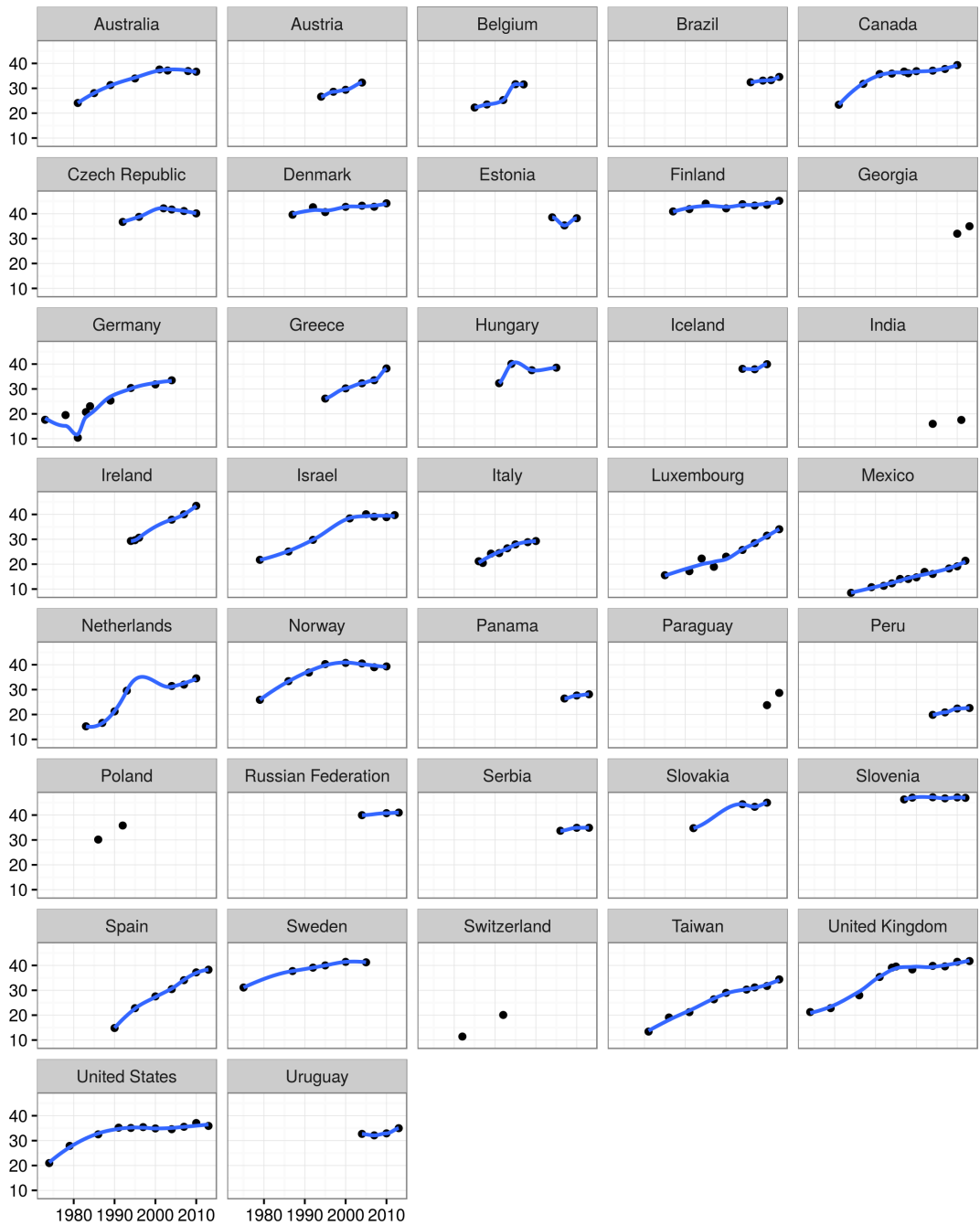

**Figure 3: Women's earning share of net household earnings (%)**

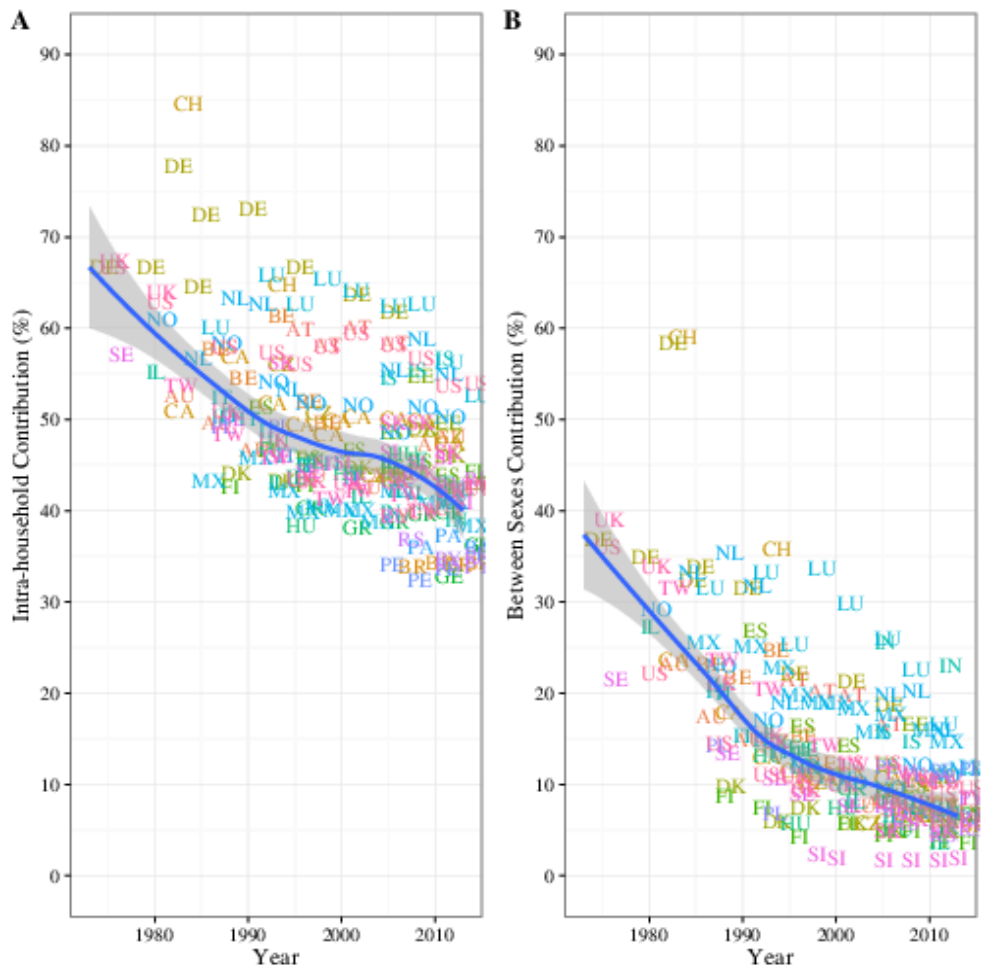

**Figure 4: Intra-household and between sex contribution to overall Theil**

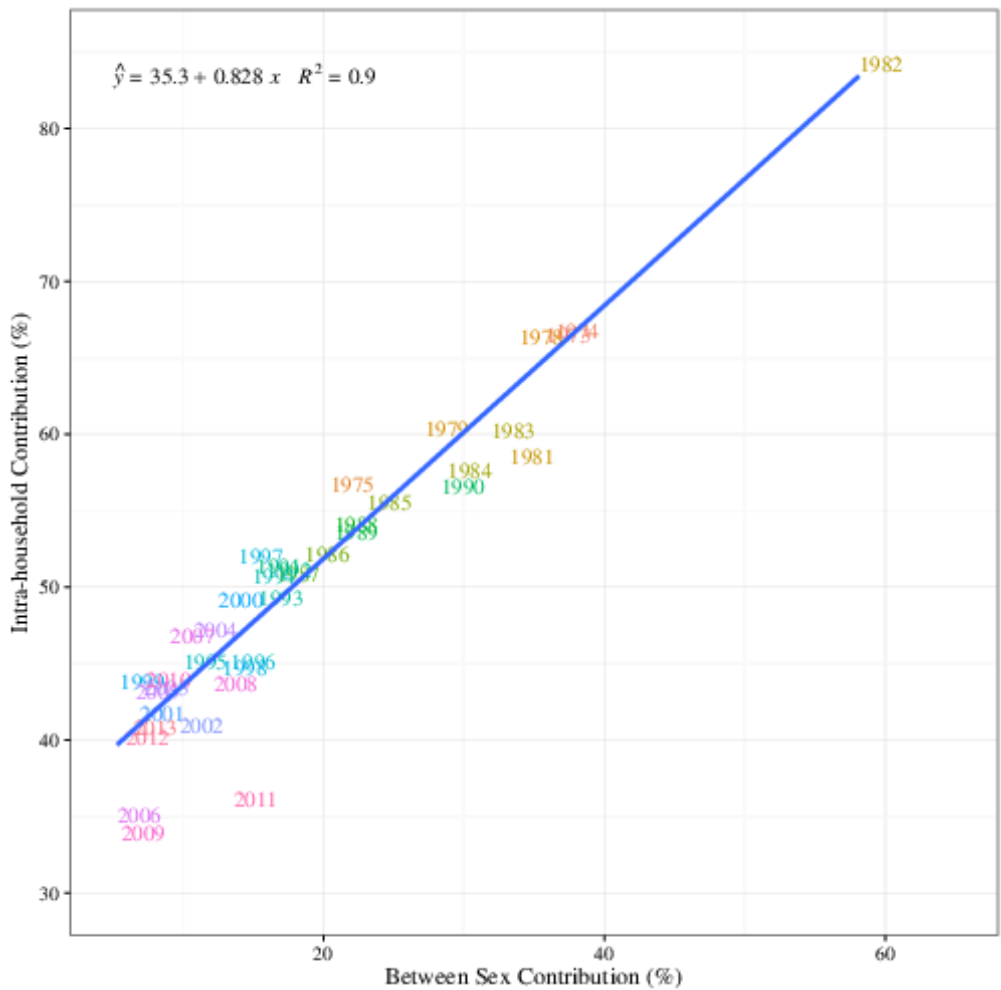

**Figure 5: Correlation between intra-household and between sex contribution**

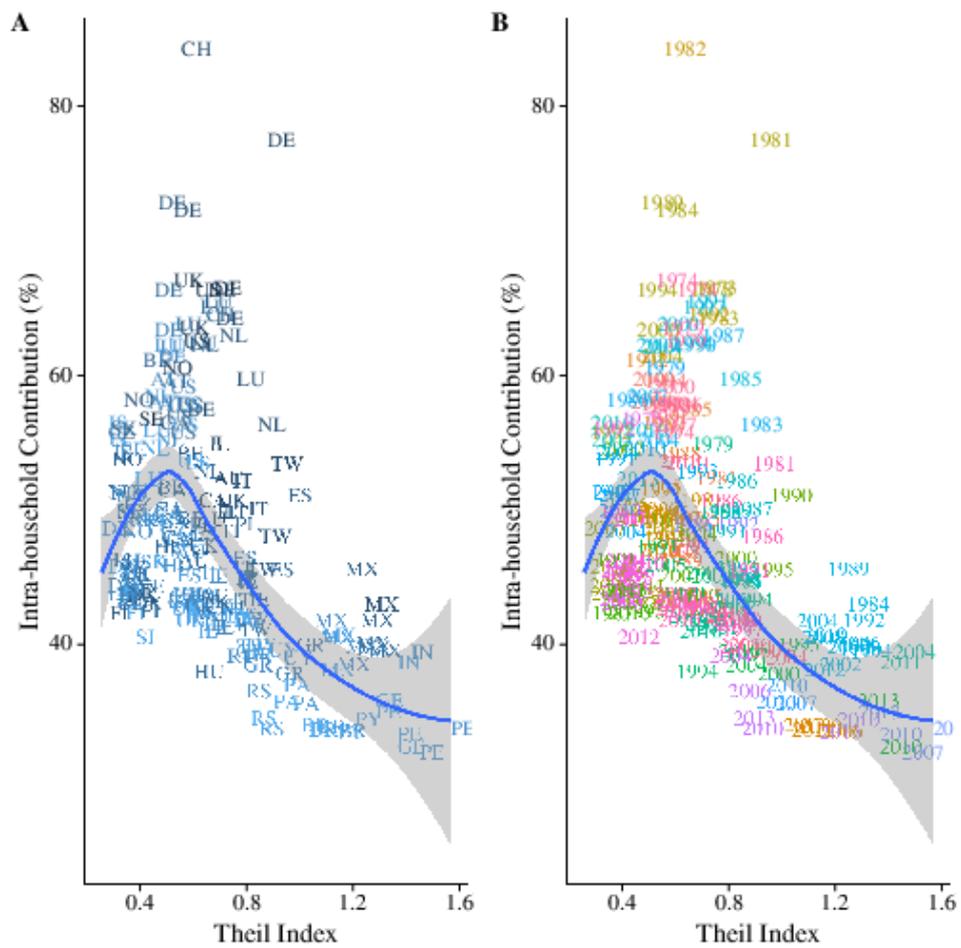

**Figure 6: Intra-household contribution and overall Theil index**

# Appendix

# The Atkinson Inequality Index as an Intra-household Welfare Loss Metric

The intra-household distribution of income $\Phi^i$ is derived from the distribution of personal incomes of *k* adults within the household:

$$\Phi^i = \Phi\big(Y_1^i, Y_2^i, \ldots, Y_j^i, \ldots, Y_{k-1}^i, Y_k^i\big) \qquad [A.1]$$

Consider an elementary additive social welfare function, W (·) defined for each household, *i* that is computed as a simple average of individual utilities, *U*, that takes individual net income $\big(Y_j^i\big)$ as the argument.

$$W_j^i = \frac{1}{k}\sum_{j=1}^{j=k} U_j^i\big(Y_j^i\big) \qquad [A.2]$$

Using Atkinson's specification (1970) for $U_j^i$

$$U_j^i(Y_j^i) = \begin{cases} \dfrac{\big(Y_j^i\big)^{1-\varepsilon_j^i}}{1-\varepsilon_j^i}; & \varepsilon_j^i \neq 1, \varepsilon_j^i \geq 0 \\ ln(Y_j^i); & \varepsilon_j^i = 1 \end{cases} \qquad [A.3]$$

The values taken by the inequality aversion parameter $\big(\varepsilon_j^i\big)$ determine the functional form of Eq. (A.3). With $\varepsilon_j^i = 0$, Eq. (A.3) reduces to a utilitarian social welfare function (SWF), consistent with perfect income pooling.

As $\varepsilon_j^i \to \infty$, Eq. (A.3) assumes the Rawlsian form. From the perspective of person *j* in household *i*, $\varepsilon$ fully characterizes the trade-offs consistent with extant intra-household distribution of income. This formulation underscores the fact that $\varepsilon$ can vary across household members.

To calculate welfare loss from intra-household income inequality, we first compute the equivalent equal income $\Theta_j^i$ following Eq. (4):

$$\frac{1}{k}\sum_{j=1}^{j=k} U_j^i\big(Y_j^i\big) = U_j^i\big(\Theta_j^i\big) = W_j^i \qquad [A.4]$$

Combining Eqs. (A.3) and (A.4),

$$\Theta_j^i = \begin{cases} \left(\frac{1}{k}\Sigma_j\big(\big(Y_j^i\big)^{1-\varepsilon_j^i}\big)\right)^{\frac{1}{1-\varepsilon_j^i}}; & \varepsilon_j^i \neq 1, \varepsilon_j^i \geq 0 \\ \big(\Pi_j(Y_j^i)^{\frac{1}{k}}\big); & \varepsilon_j^i = 1 \end{cases} \qquad [A.5]$$

The Atkinson Welfare loss metric $\Delta A_j^i$, is evaluated by substituting Eq. (A.5) in Eq. (5). For ε = 1, $\Delta A$ is the same as welfare loss calculated using a Foster welfare function based on the log-mean deviation (Sen, 1997).

**Table A1: Women's earnings share, Theil, mean contribution of within household inequality, and mean welfare loss (%), by country and year**

| Regions | Year | Mean women's earnings share | Overall Theil | Mean contribution of within hh | Mean Atkinson welfare loss | |
|---|---|---|---|---|---|---|
| | | | | | e=0.25 | e=1 |
| Austria | 1994 | 26.6 | 0.48 | 58.9 | 10.1 | 48.1 |
| Austria | 1997 | 28.6 | 0.47 | 57.3 | 9.3 | 44.1 |
| Austria | 2000 | 29.4 | 0.44 | 59.2 | 9.0 | 42.4 |
| Austria | 2004 | 32.3 | 0.43 | 57.5 | 9.0 | 42.4 |
| Australia | 1981 | 24.1 | 0.67 | 51.8 | 9.9 | 47.1 |
| Australia | 1985 | 28.1 | 0.61 | 48.7 | 9.1 | 43.3 |
| Australia | 1989 | 31.3 | 0.54 | 45.6 | 8.1 | 38.3 |
| Australia | 1995 | 33.9 | 0.62 | 43.0 | 8.3 | 39.7 |
| Australia | 2001 | 37.6 | 0.59 | 41.7 | 7.9 | 37.6 |
| Australia | 2003 | 37.2 | 0.57 | 42.9 | 8.1 | 38.8 |
| Australia | 2008 | 36.9 | 0.53 | 46.3 | 8.2 | 38.8 |
| Australia | 2010 | 36.7 | 0.52 | 47.2 | 8.2 | 39.0 |
| Belgium | 1985 | 22.3 | 0.58 | 56.9 | 10.3 | 49.5 |
| Belgium | 1988 | 23.5 | 0.54 | 53.6 | 9.1 | 43.7 |
| Belgium | 1992 | 25.3 | 0.41 | 60.5 | 8.6 | 40.9 |
| Belgium | 1995 | 31.6 | 0.46 | 51.1 | 8.0 | 37.9 |
| Belgium | 1997 | 31.6 | 0.54 | 48.8 | 8.6 | 41.3 |
| Canada | 1981 | 23.4 | 0.62 | 50.1 | 8.5 | 39.8 |
| Canada | 1987 | 31.8 | 0.48 | 55.9 | 9.1 | 42.6 |
| Canada | 1991 | 35.7 | 0.47 | 50.9 | 8.3 | 39.0 |
| Canada | 1994 | 35.9 | 0.48 | 48.5 | 8.1 | 38.2 |
| Canada | 1997 | 36.7 | 0.48 | 47.3 | 8.1 | 37.9 |
| Canada | 1998 | 36.0 | 0.45 | 49.0 | 8.0 | 37.4 |
| Canada | 2000 | 36.9 | 0.46 | 49.4 | 8.1 | 37.9 |
| Canada | 2004 | 37.1 | 0.46 | 49.2 | 8.0 | 37.1 |
| Canada | 2007 | 37.8 | 0.44 | 48.9 | 7.8 | 36.3 |
| Canada | 2010 | 39.3 | 0.46 | 46.4 | 7.8 | 36.3 |
| Germany | 1973 | 17.6 | 0.68 | 65.9 | 12.4 | 59.3 |
| Germany | 1978 | 19.5 | 0.66 | 65.7 | 12.3 | 58.8 |
| Germany | 1981 | 10.4 | 0.88 | 76.9 | 18.4 | 89.0 |
| Germany | 1983 | 20.7 | 0.68 | 63.6 | 12.4 | 59.1 |
| Germany | 1984 | 23.1 | 0.53 | 71.6 | 11.7 | 55.6 |
| Germany | 1989 | 25.3 | 0.47 | 72.2 | 10.7 | 50.4 |
| Germany | 1994 | 30.4 | 0.45 | 65.7 | 9.8 | 45.8 |
| Germany | 2000 | 31.8 | 0.46 | 62.8 | 9.3 | 43.2 |
| Germany | 2004 | 33.5 | 0.47 | 60.8 | 9.3 | 43.3 |

| Regions | Year | Mean women's earnings share | Overall Theil | Mean contribution of within household | Mean Atkinson welfare loss | |
|---|---|---|---|---|---|---|
| | | | | | e=0.25 | e=1 |
| Denmark | 1987 | 39.6 | 0.35 | 43.1 | 6.1 | 28.4 |
| Denmark | 1992 | 42.6 | 0.37 | 42.4 | 6.6 | 31.0 |
| Denmark | 1995 | 40.7 | 0.33 | 44.6 | 6.3 | 29.3 |
| Denmark | 2000 | 42.8 | 0.28 | 43.8 | 5.4 | 25.3 |
| Denmark | 2004 | 43.2 | 0.28 | 43.2 | 5.5 | 25.8 |
| Denmark | 2007 | 42.8 | 0.26 | 48.0 | 5.2 | 24.3 |
| Denmark | 2010 | 44.2 | 0.28 | 45.2 | 5.5 | 25.5 |
| Switzerland | 1982 | 11.4 | 0.56 | 83.6 | 14.1 | 66.8 |
| Switzerland | 1992 | 20.1 | 0.65 | 63.9 | 11.4 | 53.8 |
| Spain | 1990 | 14.8 | 0.95 | 50.5 | 11.9 | 57.2 |
| Spain | 1995 | 22.8 | 0.89 | 45.0 | 10.9 | 52.3 |
| Spain | 2000 | 27.6 | 0.75 | 45.8 | 10.4 | 49.7 |
| Spain | 2004 | 30.5 | 0.60 | 47.6 | 9.5 | 45.3 |
| Spain | 2007 | 34.1 | 0.54 | 44.5 | 8.5 | 40.1 |
| Spain | 2010 | 37.2 | 0.61 | 43.0 | 9.2 | 44.1 |
| Spain | 2013 | 38.3 | 0.64 | 42.1 | 9.5 | 44.7 |
| Finland | 1987 | 40.9 | 0.28 | 41.7 | 4.9 | 22.8 |
| Finland | 1991 | 41.9 | 0.30 | 45.7 | 5.7 | 26.5 |
| Finland | 1995 | 44.1 | 0.40 | 41.8 | 7.0 | 32.6 |
| Finland | 2000 | 42.2 | 0.32 | 43.3 | 6.1 | 28.4 |
| Finland | 2004 | 43.8 | 0.32 | 42.7 | 6.0 | 28.0 |
| Finland | 2007 | 43.3 | 0.33 | 43.4 | 6.4 | 29.9 |
| Finland | 2010 | 43.6 | 0.35 | 41.7 | 6.5 | 30.5 |
| Finland | 2013 | 45.2 | 0.34 | 43.4 | 6.3 | 29.1 |
| Greece | 1995 | 26.1 | 0.99 | 39.3 | 9.1 | 44.0 |
| Greece | 2000 | 30.2 | 0.91 | 37.2 | 8.6 | 41.6 |
| Greece | 2004 | 32.3 | 0.79 | 37.9 | 8.6 | 41.2 |
| Greece | 2007 | 33.5 | 0.79 | 38.6 | 8.6 | 41.1 |
| Greece | 2010 | 38.2 | 0.76 | 38.9 | 8.6 | 41.3 |
| Ireland | 1994 | 29.3 | 0.81 | 42.7 | 9.2 | 43.9 |
| Ireland | 1995 | 29.7 | 0.77 | 44.0 | 9.3 | 44.6 |
| Ireland | 1996 | 30.6 | 0.76 | 44.4 | 9.4 | 44.5 |
| Ireland | 2004 | 37.9 | 0.66 | 44.5 | 9.4 | 44.8 |
| Ireland | 2007 | 40.0 | 0.63 | 44.7 | 9.3 | 44.0 |
| Ireland | 2010 | 43.5 | 0.71 | 42.0 | 9.7 | 46.4 |
| Iceland | 2004 | 38.1 | 0.30 | 53.6 | 5.8 | 26.4 |
| Iceland | 2007 | 37.9 | 0.28 | 54.5 | 5.5 | 25.0 |
| Iceland | 2010 | 39.9 | 0.28 | 55.8 | 5.7 | 25.9 |

| Regions | Year | Mean women's earnings share | Overall Theil | Mean contribution of within household | Mean Atkinson welfare loss | |
|---|---|---|---|---|---|---|
| | | | | | e=0.25 | e=1 |
| Italy | 1986 | 21.2 | 0.75 | 51.5 | 11.1 | 53.6 |
| Italy | 1987 | 20.5 | 0.81 | 49.5 | 10.9 | 52.7 |
| Italy | 1989 | 24.2 | 0.73 | 49.2 | 10.3 | 49.8 |
| Italy | 1991 | 24.4 | 0.71 | 47.8 | 10.2 | 49.0 |
| Italy | 1993 | 26.4 | 0.75 | 45.1 | 10.0 | 48.1 |
| Italy | 1995 | 27.9 | 0.76 | 44.1 | 9.8 | 47.5 |
| Italy | 1998 | 28.9 | 0.77 | 44.2 | 9.8 | 47.0 |
| Italy | 2000 | 29.3 | 0.75 | 42.7 | 9.5 | 46.0 |
| Luxembourg | 1985 | 15.6 | 0.76 | 59.1 | 11.8 | 57.0 |
| Luxembourg | 1991 | 17.2 | 0.64 | 64.9 | 12.4 | 59.3 |
| Luxembourg | 1994 | 22.3 | 0.59 | 61.7 | 11.3 | 53.9 |
| Luxembourg | 1997 | 18.9 | 0.63 | 64.5 | 11.8 | 56.3 |
| Luxembourg | 2000 | 23.0 | 0.53 | 63.2 | 10.4 | 49.2 |
| Luxembourg | 2004 | 25.7 | 0.47 | 61.5 | 9.3 | 43.7 |
| Luxembourg | 2007 | 28.5 | 0.45 | 61.7 | 9.0 | 42.6 |
| Luxembourg | 2010 | 31.5 | 0.41 | 55.3 | 7.9 | 37.2 |
| Luxembourg | 2013 | 34.0 | 0.38 | 51.7 | 7.3 | 34.6 |
| Netherlands | 1983 | 15.3 | 0.84 | 55.7 | 11.4 | 54.7 |
| Netherlands | 1987 | 16.6 | 0.70 | 62.4 | 12.0 | 57.8 |
| Netherlands | 1990 | 21.2 | 0.59 | 61.6 | 10.5 | 50.0 |
| Netherlands | 1993 | 29.6 | 0.60 | 52.3 | 9.5 | 45.1 |
| Netherlands | 2004 | 31.5 | 0.46 | 54.5 | 8.2 | 38.3 |
| Netherlands | 2007 | 32.0 | 0.42 | 57.8 | 7.9 | 36.9 |
| Netherlands | 2010 | 34.5 | 0.41 | 54.0 | 7.6 | 35.4 |
| Norway | 1979 | 25.9 | 0.48 | 60.0 | 9.4 | 43.7 |
| Norway | 1986 | 33.4 | 0.34 | 57.5 | 6.8 | 31.4 |
| Norway | 1991 | 36.9 | 0.30 | 53.2 | 6.0 | 27.8 |
| Norway | 1995 | 40.3 | 0.28 | 50.8 | 5.6 | 26.1 |
| Norway | 2000 | 40.7 | 0.29 | 50.7 | 6.1 | 28.0 |
| Norway | 2004 | 40.5 | 0.33 | 47.7 | 6.5 | 29.7 |
| Norway | 2007 | 39.0 | 0.31 | 50.5 | 6.1 | 28.0 |
| Norway | 2010 | 39.3 | 0.31 | 49.3 | 6.0 | 27.3 |
| Sweden | 1975 | 31.1 | 0.39 | 56.2 | 7.7 | 36.0 |
| Sweden | 1987 | 37.7 | 0.31 | 49.2 | 6.4 | 29.6 |
| Sweden | 1992 | 39.1 | 0.33 | 45.1 | 6.2 | 28.7 |
| Sweden | 1995 | 40.0 | 0.36 | 43.4 | 6.5 | 30.0 |
| Sweden | 2000 | 41.4 | 0.32 | 42.5 | 5.9 | 27.1 |
| Sweden | 2005 | 41.2 | 0.31 | 42.8 | 5.9 | 27.3 |

| Regions | Year | Mean women's earnings share | Overall Theil | Mean contribution of within household | Mean Atkinson welfare loss | |
|---|---|---|---|---|---|---|
| | | | | | e=0.25 | e=1 |
| UK | 1974 | 21.3 | 0.53 | 66.5 | 10.7 | 50.3 |
| UK | 1979 | 22.8 | 0.55 | 62.9 | 10.5 | 49.3 |
| UK | 1986 | 28.0 | 0.69 | 50.1 | 9.8 | 46.6 |
| UK | 1991 | 35.3 | 0.58 | 46.7 | 8.6 | 40.7 |
| UK | 1994 | 39.1 | 0.62 | 42.6 | 8.6 | 40.7 |
| UK | 1995 | 39.5 | 0.64 | 42.2 | 8.5 | 40.5 |
| UK | 1999 | 38.4 | 0.57 | 42.2 | 7.9 | 37.1 |
| UK | 2004 | 39.8 | 0.53 | 42.3 | 7.6 | 35.8 |
| UK | 2007 | 39.6 | 0.52 | 43.1 | 7.8 | 36.7 |
| UK | 2010 | 41.5 | 0.53 | 41.2 | 7.7 | 36.3 |
| UK | 2013 | 41.8 | 0.52 | 42.0 | 7.7 | 36.2 |
| United States | 1974 | 21.0 | 0.60 | 65.8 | 11.5 | 54.3 |
| United States | 1979 | 27.9 | 0.56 | 62.0 | 10.3 | 48.6 |
| United States | 1986 | 32.5 | 0.54 | 57.1 | 9.7 | 45.4 |
| United States | 1991 | 35.2 | 0.50 | 56.3 | 8.9 | 41.7 |
| United States | 1994 | 35.1 | 0.51 | 55.2 | 8.8 | 41.1 |
| United States | 1997 | 35.4 | 0.50 | 57.1 | 8.8 | 41.3 |
| United States | 2000 | 34.9 | 0.51 | 58.5 | 9.0 | 42.2 |
| United States | 2004 | 34.6 | 0.53 | 57.2 | 9.3 | 43.6 |
| United States | 2007 | 35.6 | 0.52 | 55.7 | 9.0 | 42.1 |
| United States | 2010 | 37.1 | 0.56 | 52.9 | 9.2 | 43.5 |
| United States | 2013 | 35.9 | 0.54 | 53.0 | 9.0 | 42.5 |

| Regions | Year | Mean women's earnings share | Overall Theil | Mean contribution of within household | Mean Atkinson welfare loss | |
|---|---|---|---|---|---|---|
| | | | | | e=0.25 | e=1 |
| **Eastern Europe** | | | | | | |
| Czech Republic | 1992 | 36.7 | 0.28 | 55.1 | 5.9 | 27.9 |
| Czech Republic | 1996 | 38.8 | 0.34 | 49.8 | 6.6 | 31.1 |
| Czech Republic | 2002 | 42.2 | 0.51 | 42.9 | 7.9 | 37.8 |
| Czech Republic | 2004 | 41.7 | 0.50 | 42.5 | 7.9 | 37.7 |
| Czech Republic | 2007 | 41.1 | 0.48 | 47.9 | 8.3 | 39.8 |
| Czech Republic | 2010 | 40.1 | 0.50 | 47.3 | 8.4 | 40.2 |
| Estonia | 2004 | 38.6 | 0.40 | 43.7 | 6.7 | 31.2 |
| Estonia | 2007 | 35.3 | 0.32 | 53.9 | 6.3 | 29.1 |
| Estonia | 2010 | 38.2 | 0.42 | 48.7 | 7.6 | 35.5 |
| Hungary | 1991 | 32.3 | 0.45 | 46.6 | 8.0 | 37.6 |
| Hungary | 1994 | 40.1 | 0.60 | 37.4 | 8.7 | 41.7 |
| Hungary | 1999 | 37.5 | 0.55 | 43.1 | 8.9 | 42.3 |
| Hungary | 2005 | 38.6 | 0.48 | 45.3 | 8.4 | 39.8 |
| Poland | 1986 | 30.2 | 0.59 | 48.3 | 8.7 | 42.0 |
| Poland | 1992 | 35.8 | 0.76 | 48.3 | 10.8 | 52.2 |
| Serbia | 2006 | 33.7 | 0.80 | 35.9 | 8.7 | 42.1 |
| Serbia | 2010 | 34.9 | 0.85 | 33.2 | 8.3 | 40.0 |
| Serbia | 2013 | 34.9 | 0.82 | 33.9 | 8.3 | 40.1 |
| Russia | 2004 | 40.0 | 0.73 | 38.6 | 9.3 | 44.0 |
| Russia | 2010 | 40.8 | 0.57 | 41.9 | 8.4 | 39.9 |
| Russia | 2013 | 41.0 | 0.53 | 42.6 | 7.9 | 37.5 |
| Slovenia | 1997 | 46.3 | 0.34 | 44.2 | 6.4 | 30.4 |
| Slovenia | 1999 | 47.0 | 0.36 | 44.5 | 7.0 | 33.5 |
| Slovenia | 2004 | 47.1 | 0.33 | 45.6 | 6.6 | 31.5 |
| Slovenia | 2007 | 46.7 | 0.32 | 44.2 | 6.5 | 30.8 |
| Slovenia | 2010 | 47.1 | 0.32 | 44.9 | 6.5 | 30.8 |
| Slovenia | 2012 | 46.9 | 0.38 | 40.0 | 7.0 | 33.0 |
| Slovakia | 1992 | 34.7 | 0.29 | 55.4 | 6.5 | 30.6 |
| Slovakia | 2004 | 44.4 | 0.37 | 48.6 | 6.9 | 32.7 |
| Slovakia | 2007 | 43.3 | 0.33 | 48.8 | 6.6 | 31.5 |
| Slovakia | 2010 | 45.0 | 0.40 | 45.5 | 7.6 | 36.1 |

| Regions | Year | Mean women's earnings share | Overall Theil | Mean contribution of within household | Mean Atkinson welfare loss | |
|---|---|---|---|---|---|---|
| | | | | | e=0.25 | e=1 |
| **South and Central Americas** | | | | | | |
| Brazil | 2006 | 32.4 | 1.15 | 33.0 | 9.8 | 47.0 |
| Brazil | 2009 | 33.1 | 1.07 | 33.3 | 9.7 | 46.4 |
| Brazil | 2011 | 33.3 | 1.04 | 33.1 | 9.5 | 45.5 |
| Brazil | 2013 | 34.6 | 1.00 | 33.4 | 9.4 | 45.4 |
| Mexico | 1984 | 8.5 | 1.24 | 42.4 | 12.0 | 57.9 |
| Mexico | 1989 | 10.8 | 1.17 | 45.0 | 12.8 | 61.7 |
| Mexico | 1992 | 11.3 | 1.23 | 41.2 | 12.3 | 59.5 |
| Mexico | 1994 | 12.3 | 1.26 | 38.9 | 11.9 | 57.3 |
| Mexico | 1996 | 14.1 | 1.21 | 39.5 | 11.7 | 56.5 |
| Mexico | 1998 | 14.0 | 1.21 | 39.1 | 11.8 | 56.9 |
| Mexico | 2000 | 14.7 | 1.17 | 39.2 | 11.7 | 56.5 |
| Mexico | 2002 | 16.9 | 1.14 | 37.9 | 11.5 | 55.5 |
| Mexico | 2004 | 16.1 | 1.05 | 41.2 | 12.3 | 59.1 |
| Mexico | 2008 | 18.3 | 1.08 | 39.9 | 11.7 | 56.2 |
| Mexico | 2010 | 19.1 | 1.09 | 40.1 | 11.8 | 56.6 |
| Mexico | 2012 | 21.4 | 1.08 | 37.5 | 11.3 | 54.5 |
| Panama | 2007 | 26.4 | 0.97 | 35.0 | 10.9 | 52.2 |
| Panama | 2010 | 27.6 | 0.94 | 36.4 | 11.0 | 52.7 |
| Panama | 2013 | 28.1 | 0.90 | 35.1 | 10.8 | 51.9 |
| Peru | 2004 | 19.9 | 1.57 | 33.2 | 8.7 | 41.8 |
| Peru | 2007 | 20.9 | 1.45 | 31.4 | 9.1 | 43.7 |
| Peru | 2010 | 22.4 | 1.37 | 32.7 | 9.2 | 44.3 |
| Peru | 2013 | 22.6 | 1.29 | 34.4 | 9.3 | 44.6 |
| Paraguay | 2010 | 23.8 | 1.21 | 33.8 | 8.6 | 41.3 |
| Paraguay | 2013 | 28.7 | 1.14 | 32.9 | 8.9 | 42.7 |
| Uruguay | 2004 | 32.7 | 0.94 | 38.4 | 10.6 | 51.1 |
| Uruguay | 2007 | 32.1 | 0.88 | 38.9 | 10.3 | 49.4 |
| Uruguay | 2010 | 32.9 | 0.80 | 39.4 | 10.1 | 48.3 |
| Uruguay | 2013 | 35.0 | 0.71 | 41.5 | 9.8 | 46.9 |

|  |  |  |  | Mean |  |  |
|---|---|---|---|---|---|---|
| Regions | Year | Mean women's earnings share | Overall Theil | contribution of within household | Mean Atkinson welfare loss | |
|  |  |  |  |  | e=0.25 | e=1 |
| **Asia** |  |  |  |  |  |  |
| Georgia | 2010 | 32.0 | 1.37 | 31.8 | 9.1 | 43.6 |
| Georgia | 2013 | 35.0 | 1.28 | 35.2 | 10.1 | 48.3 |
| India | 2004 | 16.0 | 1.42 | 38.8 | 10.0 | 48.1 |
| India | 2011 | 17.5 | 1.37 | 38.0 | 10.4 | 49.6 |
| Taiwan | 1981 | 13.4 | 0.89 | 52.8 | 11.8 | 56.8 |
| Taiwan | 1986 | 19.1 | 0.85 | 47.4 | 10.2 | 49.2 |
| Taiwan | 1991 | 21.2 | 0.80 | 44.9 | 9.6 | 46.3 |
| Taiwan | 1997 | 26.4 | 0.77 | 40.4 | 8.9 | 42.9 |
| Taiwan | 2000 | 29.0 | 0.76 | 41.3 | 9.0 | 43.3 |
| Taiwan | 2005 | 30.3 | 0.75 | 40.9 | 9.0 | 43.4 |
| Taiwan | 2007 | 31.2 | 0.76 | 39.2 | 8.9 | 42.9 |
| Taiwan | 2010 | 31.7 | 0.74 | 41.2 | 9.2 | 44.1 |
| Taiwan | 2013 | 34.4 | 0.72 | 41.6 | 9.0 | 43.5 |
| **Middle East** |  |  |  |  |  |  |
| Israel | 1979 | 21.8 | 0.66 | 54.3 | 10.8 | 51.5 |
| Israel | 1986 | 25.1 | 0.69 | 49.3 | 10.5 | 50.0 |
| Israel | 1992 | 29.8 | 0.74 | 42.2 | 9.9 | 47.0 |
| Israel | 2001 | 38.4 | 0.68 | 40.7 | 9.0 | 42.7 |
| Israel | 2005 | 40.1 | 0.67 | 41.2 | 9.2 | 43.9 |
| Israel | 2007 | 39.0 | 0.64 | 40.9 | 9.1 | 43.3 |
| Israel | 2010 | 38.9 | 0.62 | 40.4 | 9.0 | 42.8 |
| Israel | 2012 | 39.7 | 0.57 | 41.5 | 8.6 | 40.8 |